\documentclass[aps,pra,onecolumn,11pt]{revtex4-1}
\usepackage[dvipdfm]{graphicx}
\usepackage[usenames,dvipsnames]{xcolor}
\usepackage[dvipdfm]{hyperref}
\usepackage{amsmath}
\usepackage{graphicx}
\usepackage{subcaption}
\begin{document}

\title{Quantized damped transversal single particle mechanical waves}
\author{Ferenc M\'arkus}
	\email[Corresponding author: ]{markus.ferenc@ttk.bme.hu; markus@phy.bme.hu}
	\affiliation{Department of Physics,
		Budapest University of Technology and Economics,
		M\H{u}egyetem rkp. 3, H-1111 Budapest, Hungary}

	\author{Katalin Gamb\'ar}
	\email[ ]{gambar.katalin@uni-obuda.hu}
	\affiliation{Department of Natural Sciences, Institute of Electrophysics, \\ Kálmán Kandó Faculty of Electrical Engineering, \\ Óbuda University, \\ Tavaszmező u. 17, H-1084 Budapest, Hungary}
    
    \affiliation{Department of Natural Sciences, \\ National University of Public Service, \\  Ludovika t\'er 2, H-1083 Budapest, Hungary}

\date{\today}

\begin{abstract}
In information transfer, the dissipation of a signal may have crucial importance. The feasibility of reconstructing the distorted signal also depends on this. That is why the study of quantized dissipative transversal single particle mechanical waves may have an important role. It may be true, particularly on the nanoscale in the case of signal distortion, loss, or restoration. Based on the damped oscillator quantum description, we generalize the canonical quantization procedure for the transversal waves. Furthermore, we deduce the related damped wave equation and the state function. We point out the two kinds of solutions of the wave equation. One involves the well-known spreading solution superposed with the oscillation, in which the loss of information is complete. The other is the Airy function solution, which is non-spreading, so there is information loss only due to oscillation damping. However, the structure of the wavefront remains unchanged. Thus, this result allows signal reconstruction, which is important in restoring the lost information. 
\end{abstract}

\maketitle

\section{Introduction}

The quantum-based signal propagation is the focus of information transfer today. The correct description of the signal loss due to the dissipation or the possible restoration after the loss effect is a challenge in the examination. It seems obvious if we knew the correct formulation of the mechanism of loss, the reconstruction of the signal may be available. To achieve this aim seems a very long way. In the recent work, we present the quantum mechanical problem of the damped transversal wave propagation. This mechanic process interacting with other physical phenomena, like thermal energy \cite{markus2021}, electric charge transfers, or even spin waves, spin relaxation processes \cite{bence_markus_2020,bence_markus_2023,csosz2020,csosz2020_2}, in spin wave computing \cite{mahmoud2020} may cause unexpected deviance in their realization or in the signal reconstruction. Further experimental and theoretical motivations to integrate consequent the quantum dissipation in the quantum theories are in Haken's \cite{haken1970} and Haake's \cite{haake1973} early works. \\
The wish to develop a consequent quantization procedure of the damped oscillator goes back to the early beginnings of quantum theory. The ideas and methods differ like based on non-conserving subsystems (Caldirola, and Kanai) \cite{caldirola1941,kanai1948}, explicit time-dependent formulation (Dekker, and Dittrich {\it el al.}) \cite{dekker1981,dittrich1996} that brings an incompatibility with the uncertainty principle (Weiss) \cite{weiss2012}, decoupled harmonic oscillators in a reservoir (Leggett, and Caldeira {\it el al.}) \cite{leggett1984,caldeira1993,rosenau2000}, or by developing Caldirola's, and Kanai's procedure (Choi) \cite{choi2013}. It is pointed out (Bagarello {\it el al.}) \cite{bagarello2019}, why it is impossible to quantize the Bateman's damped oscillator description \cite{bateman1931}. A second quantized solution is presented by Risken \cite{risken1989}. In this method there is no frequency shift, just the amplitude is damping. To study the damped quantum oscillator a WKB approximation-based quantization procedure is built up by Serhan et al. \cite{serhan2018,serhan2019}. It is an explicit time-dependent Lagrangian method resulting exponentially decreasing time-dependent wave function, but it is a standing solution in space. For quantum dissipative systems El-Nabulsi \cite{elnabulsi2020} supposed the introduction a dynamical friction in the theory. However, the velocity-dependent term does not appear in the equation of motion, which generates the exponential time relaxation. \\
We build up the transversal wave from a local vertical damped oscillation and a free propagation along horizontal direction. The internal shearing causes the damping effect, and the generated longitudinal motion is considered dissipation-free.

\section{The canonical description of a damped harmonic oscillator}  \label{Sec_Lagrangian}

In the transversal propagation a shear interaction appears. We assume that this shearing effect is
proportional to the oscillation velocity, and its $y$ direction of motion is perpendicular to the $x$ direction of propagation. The shearing makes the oscillator motion dissipative. \\
We divide the calculations into parts. We examine first the Lagrangian description of the motion of the damped harmonic oscillator in the elaboration 
\begin{equation}
\ddot y + 2\lambda \dot y + \omega^2 y = 0,
\end{equation}
where $m$ is the mass of unit length of the string, $\lambda$ is a specific damping factor, and $\omega$ is the angular frequency of the free oscillator. The Lagrangian formulation requires the introduction of a generator potentials in the cases of non-selfadjoint operators \cite{gambar1994,gambar2020}. From this reason we need to express the measurable quantity $y$ by the potential $q$ as the definition equation shows \cite{szegleti2020}, 
\begin{equation}
y = \ddot q - 2\lambda \dot q + \omega^2 q.  \label{potencial_definition}
\end{equation}
Then, we can construct a potential based Lagrangian that can serve the above equation of motion
\begin{equation}
L = \frac{1}{2} \left( \ddot q - 2\lambda \dot q + \omega^2 q \right)^2.
\end{equation}
In the case of second order derivatives there are two generalized coordinates and two canonical momenta \cite{markus_gambar_2022}. Applying the related calculation procedure, we obtain that the coordinates are
\begin{equation}
q_1 := q,  \label{coordinate_q_1}
\end{equation}
and
\begin{equation}
q_2 := \dot{q}.  \label{coordinate_q_2}
\end{equation}
In this particular case the conjugated momentum $p_1$ to $q_1$ is
\begin{equation}
p_1 := 4 \lambda^2 \dot{q} - \dddot{q\hspace{0pt}} - \omega^2 \dot{q} -2 \lambda \omega^2 q.  \label{momentum_p_1}
\end{equation}
The second momentum $p_2$ to $q_2$ is
\begin{equation}
p_2 := \ddot{q} - 2 \lambda \dot{q} + \omega^2 q.  \label{momentum_p_2}
\end{equation}
In the development of the theory we need the formulation of the Hamiltonian. We can express it by the introduced canonical variables 
\begin{align}
H = p_1 \dot{q}_1 + p_2 \dot{q}_2 - L  \nonumber \\ = \frac{1}{2} p_2^2 - \omega^2 p_2 q_1 + p_1 q_2 + 2 \lambda p_2 q_2.  \label{oscillator_hamiltonian_by_momenta}
\end{align}
We obtain new coordinates $Q_1$, and $Q_2$, and new momenta $P_1$, and $P_2$, preserving the energy unit of Hamiltonian by the following transformations 
\begin{eqnarray}
P_2 &=& m \omega p_2,  \label{P_2}  \\
Q_2 &=& \omega q_2,  \label{Q_2}  \\
P_1 &=& m \omega p_1,  \label{P_1}  \\
Q_1 &=& \omega q_1.  \label{Q_1}
\end{eqnarray}
The canonical momentum $P_2$ and coordinate $Q_2$ become the usual mechanical momentum and spatial coordinate. The transformation of the Hamiltonian is  
\begin{equation}
H' = m \omega^2 H.  \label{H'}
\end{equation}
The units of the quantities are denoted by the bracket $[ \quad ]$. Finally, the Hamiltonian of the damped oscillator is
\begin{equation}
H' = \frac{1}{2m} P_2^2 - \omega^2 P_2 Q_1 + P_1 Q_2 + 2 \lambda P_2 Q_2.  \label{oscillator_hamiltonian_by_new_momenta}
\end{equation}

Applying the expressions of coordinates in Eqs. (\ref{coordinate_q_1}) and (\ref{coordinate_q_2}), and momenta (\ref{momentum_p_1}) and (\ref{momentum_p_2}), the calculation of the Hamiltonian results \cite{markus_gambar_2022} that 
\begin{equation}
H = 0.  
\end{equation}
Due to the proportionality of $H$ and $H'$ in Eq. (\ref{H'}) it is apparent that 
\begin{equation}
H'=0  \label{zero-valued_Hamiltonian}
\end{equation} 
in Eq. (\ref{oscillator_hamiltonian_by_new_momenta}). Since the treated system is dissipative and the Lagrangian has no explicit time dependence, the only possibility for the constant Hamiltonian neeeds to be zero. In this way there is no contradiction in the theory \cite{markus_gambar_2022}.

\section{The quantization of the damped transversal single particle mechanical wave}  \label{damped_wave_quantization}

We can arrive at the state equation of the quantized damped wave identifying the canonical momenta in the Hamiltonian, $H'$, in Eq. (\ref{oscillator_hamiltonian_by_new_momenta}). Due to the transversal wave motion, the direction of wave propagation along the axis x and the the displacement in the direction of y, the canonical pair $(P_2,Q_2)$ relates two-dimensional variables: $(P_{2x},P_{2y})$ and $(Q_{2x},Q_{2y})$. The relevant forms come from the Eqs. (\ref{P_2}) and (\ref{Q_2}). The damped oscillator motion is in the direction of y, thus the transformed conjugated pair of the spatial coordinate $y$ is as usual
\begin{equation}
P_{2y} \left( = mv_y = m\frac{\text{d}y}{{\mathrm{d}}t} = \right) = \hbar k_y , \,\,\,\,\,\,\,\,\,\, Q_{2y} = y .  \label{P_2y_Q_2y}
\end{equation}
However, in spite of the free wave propagation in the direction of x there is no actual displacement in this direction. We consider that the torsion effect caused a contribution to the longitudinal $P_{2x}$ mechanical momentum. So, we may obtain the related canonical variables    
\begin{equation}
P_{2x} \left( = mv_x = m\frac{\text{d}x}{{\mathrm{d}}t} = \right) = \hbar k_x , \,\,\,\,\,\,\,\,\,\, Q_{2x} = x = 0 .  \label{P_2x_Q_2x}
\end{equation}
The construction of the momentum $P_1 = (P_{1x},P_{1y})$ and the coordinate $Q_1 = (Q_{1x},Q_{1y})$ are based on the Eqs. (\ref{P_1}) and (\ref{Q_1}), and a comparison with the momentum $P_2$ and the coordinate $Q_2$ in Eqs. (\ref{P_2y_Q_2y}) and (\ref{P_2x_Q_2x}). The appearing time factor in Eqs. (\ref{P_1}) and (\ref{Q_1}) can be associated with Fourier transformed pairs, i.e.,   
\begin{equation}
P_{1y} = -\mathrm{i} \hbar k_y \omega, \,\,\,\,\,\,\,\,\,\, Q_{1y} = \frac{y}{\mathrm{i}\omega},   \label{P_1y_Q_1y}
\end{equation}
and
\begin{equation}
P_{1x} = -\mathrm{i} \hbar k_x \omega, \,\,\,\,\,\,\,\,\,\, Q_{1x} = \frac{x}{\mathrm{i}\omega} = 0.  \label{P_1x_Q_1x}
\end{equation}
To introduce the operator calculus, we need the relevant Fourier transforms 
\begin{flalign}
&\frac{\partial}{\partial y} = \mathrm{i}k_y, \,\,\,\,\,\,\,\,\,\, \frac{\partial}{\partial x} = \mathrm{i}k_x, \nonumber \\ \frac{\partial}{\partial t} = -\mathrm{i}\omega, \,\,\,\,\,\,\,\,\,\, &\frac{1}{-\mathrm{i}\omega} = \int ... \,\, {\mathrm{d}}t, \,\,\,\,\,\,\,\,\,\, k_y y \longrightarrow \frac{1}{\mathrm{i}}.  \label{Fourier_terms}
\end{flalign}

\subsection{Commutation rules}

Applying the above Fourier formulations, we can get the following commutation rules
\begin{equation}
\left[ P_{2y},Q_{2y} \right] = \frac{\hbar}{\mathrm{i}}, \,\,\,\,\,\,\,\,\,\, \left[ P_{2x},Q_{2x} \right] = 0.
\end{equation}
\begin{equation}
\left[ P_{1y},Q_{1y} \right] = - \frac{\hbar}{\mathrm{i}}, \,\,\,\,\,\,\,\,\,\, \left[ P_{1x},Q_{1x} \right] = 0,
\end{equation}
\begin{equation}
\left[ P_{2y},Q_{1y} \right] = 0, \,\,\,\,\,\,\,\,\,\, \left[ P_{2x},Q_{1x} \right] = 0,
\end{equation}
\begin{equation}
\left[ P_{1y},Q_{2y} \right] = 0, \,\,\,\,\,\,\,\,\,\, \left[ P_{1x},Q_{2x} \right] = 0.
\end{equation}
As it can be seen, the structure of the brackets reflects the usual trend.

\subsection{Calculating the terms of Hamiltonian}

Applying the above rules, in Eq. (\ref{oscillator_hamiltonian_by_new_momenta}), the terms of the Hamiltonian  can be expressed in the operator formulation. The calculation of $P_{2y}$ and its square $P_{2y}^2$ come from Eq. (\ref{P_2y_Q_2y}) and the first term in Eq. (\ref{Fourier_terms}). Similarly, $P_{2x}$ and its square $P_{2x}^2$ from Eq. (\ref{P_2x_Q_2x}), and the second term in Eq. (\ref{Fourier_terms}). Finally, we obtain the components of momentum $P_2$ and their squares
\begin{equation}
P_{2y} = \hbar k = \frac{\hbar}{\mathrm{i}}\frac{\partial}{\partial y} \,\,\,\,\, \longrightarrow \,\,\,\,\, P_{2y}^2 = - {\hbar^2}\frac{\partial^2}{\partial y^2} ,
\end{equation}
\begin{equation}
P_{2x} = \hbar k = \frac{\hbar}{\mathrm{i}}\frac{\partial}{\partial x} \,\,\,\,\, \longrightarrow \,\,\,\,\, P_{2x}^2 = - {\hbar^2}\frac{\partial^2}{\partial x^2} .
\end{equation}
Summarizing these results, we can calculate the terms of the Hamiltonian. So, the first term is
\begin{equation}
P_{2}^2 = - {\hbar^2}\frac{\partial^2}{\partial y^2} - {\hbar^2}\frac{\partial^2}{\partial x^2}.
\end{equation}
The second term includes the sum of $P_2 Q_1 = P_{2y} Q_{1y} + P_{2x} Q_{1x}$ quadratical products. Now, we consider $P_2$ from Eqs. (\ref{P_2y_Q_2y}) and (\ref{P_2x_Q_2x}), furthermore, $Q_1$ from Eqs. (\ref{P_1y_Q_1y}) and (\ref{P_1x_Q_1x}). We must take into account that $Q_{1x} = 0$. It reduces the number of the calculations. After all, we apply the fourth Fourier transform in Eq. (\ref{Fourier_terms}). The deduction steps of the second term can be followed 
\begin{equation}
P_2 Q_1 = P_{2y} Q_{1y} = \hbar k \frac{y}{\mathrm{i}\omega} = \underbrace{\frac{1}{\mathrm{i}\omega}}_{-\int ... \,\, \text{d}t} \underbrace{\hbar k}_{mv_y} y = -\int my \frac{{\mathrm{d}}y}{{\mathrm{d}}t} \,\, {\mathrm{d}}t = -\frac{1}{2} m y^2.
\end{equation}
We can recognize that only the vertical displacement has contribution to the motion. The third term of the Hamiltonian is $P_1 Q_2 = P_{1y} Q_{2y} + P_{1x} Q_{2x}$. We take $P_1$ from Eqs. (\ref{P_1y_Q_1y}) and (\ref{P_1x_Q_1x}), the components of $Q_2$ from Eqs. (\ref{P_2y_Q_2y}) and (\ref{P_2x_Q_2x}) by the restriction $Q_{2x} = 0$. Applying the third and fifth Fourier transform, we obtain 
\begin{equation}
P_1 Q_2 = P_{1y} Q_{2y} = -\mathrm{i} \hbar k \omega y = -\mathrm{i} \hbar \underbrace{k y}_{1/\mathrm{i}} \underbrace{\omega}_{\mathrm{i} \frac{\partial}{\partial t}}  = \frac{\hbar}{\mathrm{i}} \frac{\partial}{\partial t}.
\end{equation}
Finally, we focus on the fourth term of the Hamiltonian, the $P_2 Q_2$. The necessary formulations come from Eqs. (\ref{P_2y_Q_2y}) and (\ref{P_2x_Q_2x}), and we take into account the fifth Fourier transform    
\begin{equation}
P_2 Q_2 = P_{2y} Q_{2y} + P_{2x} Q_{2x} = \hbar k y = \hbar \underbrace{k y}_{1/\mathrm{i}} = \frac{\hbar}{\mathrm{i}}.
\end{equation}
After these calculation we obtain the last term of the Hamiltonian
\begin{equation}
2 \lambda P_2 Q_2 = 2 \lambda \frac{\hbar}{\mathrm{i}}.
\end{equation}

\subsection{The state equation of the damping traveling quantum wave}

Now, we turn back to the examination of the Hamiltonian. We saw that the deduced Hamiltonian $H'$ in Eq. (\ref{oscillator_hamiltonian_by_new_momenta}) must be zero as is in Eq. (\ref{zero-valued_Hamiltonian}). We substitute the above calculated expressions, and we arrive at the quantized state equation of the damping travelling mechanical wave
%
%
\begin{equation}
0  = \lefteqn{\underbrace{\phantom{- \frac{\hbar^2}{2m}\frac{\partial^2 \psi}{\partial y^2} + \frac{1}{2} m \omega^2 y^2 \psi \,\, + \frac{\hbar}{\mathrm{i}} \frac{\partial \psi}{\partial t}}}_{\text{frictionless quantum oscillator}}}{- \frac{\hbar^2}{2m}\frac{\partial^2 \psi}{\partial y^2} + \frac{1}{2} m \omega^2 y^2 \psi \,\, +}
\overbrace{\, \frac{\hbar}{\mathrm{i}} \frac{\partial \psi}{\partial t} - \, \frac{\hbar^2}{2m}\frac{\partial^2 \psi}{\partial x^2}}^{\text{free motion}} \, +
\underbrace{\,\,\,\, 2 \lambda \frac{\hbar}{\mathrm{i}} \psi \,\,\,\,}_{\text{damping}}. \label{dissipative_state_equation}
\end{equation} 
The essential effect of the zero-valued Hamiltonian can be read out from the equation. The first three terms pertain to the frictionless oscillator motion. The third and fourth terms has a role in the travelling along the axis x. It is really a quantum state equation in which the non-hermitian fifth term generates the dissipation in the system. Initially, we started our study with the damped oscillator. We will see from the solution that the damping is not restricted for the oscillation, but the entire motion involves it. The damping is unique for the vibration and the travelling motion at once. \\
The above state equation in Eq. (\ref{dissipative_state_equation}) belongs to the big family of the complex quantum potentials \cite{Razavy2005,markus2016}. These equations describe dissipative processes. It means that the probability meaning of the state function (wave function) is not valid any more.

\section{Oscillating-Travelling damped wave packet}

\subsection{The oscillating part}

Razi and Naqvi \cite{Razi2000} calculated the shape-dependent solution of the nondamped quantum harmonic oscillator. Applying the potential based canonical quantization procedure \cite{markus_gambar_2022}, the damped oscillator wave packet is
\begin{equation}
\rho(y,t) = | \Psi (y,t) |^2 = \frac{1}{\sigma_{y}(t)\sqrt{2}} \exp{ \left\{ -\frac{\left[y - y_{0} \cos(\omega t) \right]^2}{2\sigma_{y}^{2}(y,t)} \right\}} \times \exp \left( -4\lambda t \right),  
\end{equation}
where
\begin{equation}
\sigma_{y}^{2}(y,t) = \frac{\hbar}{2{\gamma}m{\omega}} \left[ \cos^2(\omega t) + {\gamma}^2 \sin^2(\omega t) \right]
\end{equation}
This oscillation is superposed to a travelling propagation. The travelling wave solution pertains to the free motion part of Eq. (\ref{dissipative_state_equation}).

\subsection{The Gaussian spreading solution of the free propagation part}

The free propagation state equation --- the overbraced part in Eq. (\ref{dissipative_state_equation}) --- is
\begin{equation}
0 = \frac{\hbar}{\mathrm{i}} \frac{\partial \psi}{\partial t} - \frac{\hbar^2}{2m}\frac{\partial^2 \psi}{\partial x^2}. \label{free_propagation_schrodinger}
\end{equation}
Its solution is a packet of velocty $v$ along the direction of x
\begin{equation}
\psi_{G}(x,t) = \frac{(4/\pi)^{1/4}}{\sqrt{a + \mathrm{i}\frac{4\hbar}{m a}t}}\exp \left( -\frac{2(x - v_x t)^2}{a^2 + \mathrm{i}\frac{4\hbar}{m}t} \right),  \label{free_propagation_wave_function}
\end{equation}
where the normalized Gaussian initial state funtion is
\begin{equation}
\psi_{G}(x,0) = \frac{(4/\pi)^{1/4}}{\sqrt{a}}\exp \left( -\frac{2x^2}{a^2} \right).  \label{initial_state_funtion}
\end{equation}
Thus the travelling wave packet contribution to the motion is
\begin{equation}
\varrho(x,t) = | \Psi (x,t) |^2 = \frac{(4/\pi)^{1/2}}{\sqrt{a^2 + \frac{16\hbar^2}{m^2 a^2}t^2}} \times \exp \left( -\frac{4(x - v_x t)^2}{a^2 + \frac{16\hbar^2}{m^2 a^2}t^2} \right).  \label{free_propagation_wave_packet}
\end{equation}
Finally, we can build up the state function of the quantized damped mechanical wave in Eq. (\ref{quantized_damped_mechanical_wave_packet}). The damping effect relates to the shearing in the direction of y. During the free propagation a spreading of the wave packet appears. The product of the $|\Psi(y,x,t)|^2 = \rho(y,t)\varrho(x,t)$ expresses the time evolution of state function of the damped transversal quantum mechanical wave
\begin{eqnarray}
|\Psi(y,x,t)|^2 &=& \frac{1}{\sigma_{y}(t)\sqrt{2}} \exp{ \left\{ -\frac{\left[y - y_{0} \cos(\omega t) \right]^2}{2\sigma_{y}^{2}(y,t)} \right\}} \times \exp \left( -4\lambda t \right) \nonumber \\  &\times& \frac{(4/\pi)^{1/2}}{\sqrt{a^2 + \frac{16\hbar^2}{m^2 a^2}t^2}} \times \exp \left( -\frac{4(x - v_x t)^2}{a^2 + \frac{16\hbar^2}{m^2 a^2}t^2} \right).  \label{quantized_damped_mechanical_wave_packet}
\end{eqnarray}
There is only one damping in the motion related to the $\exp \left( -4\lambda t \right)$ factor. Now, we can the recognize the previously mentioned situation that the damping is of the entire solution own. We do not need to, or we cannot, divide the damping to the damping of oscillation or travelling. The spreading out and the damping of the initial wave form lead to the total loss of the information. Consequently, we cannot recover the initial physical state.

\subsection{The Airy non-spreading solution of the free propagation part}

\begin{figure}[h!]
\centerline{
\includegraphics[width=0.6\columnwidth]{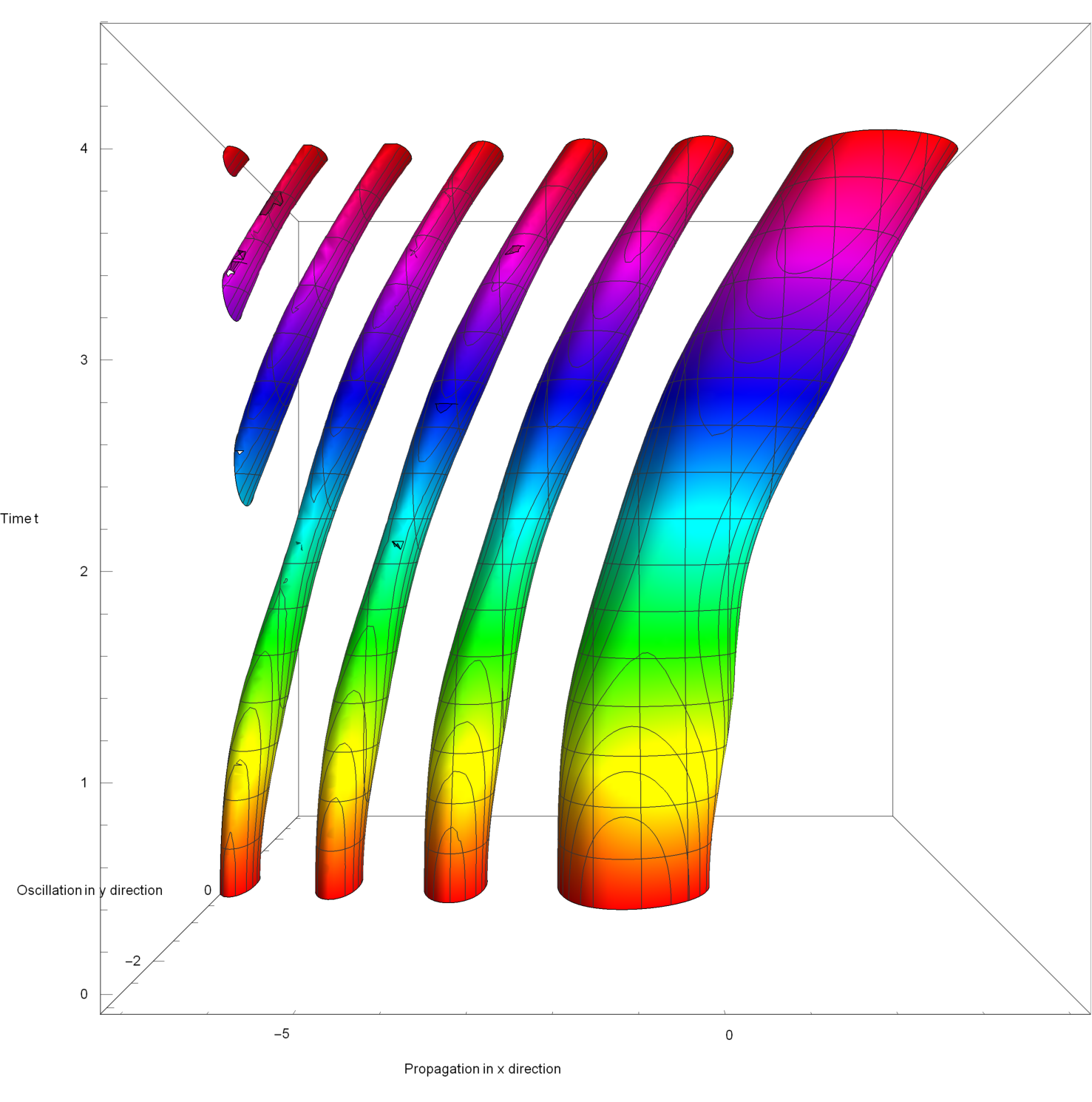}}
\caption{Nondamped oscillating-travelling Airy wave from front view. The translation motion is in the x direction, the oscillation is in the y direction. The third coordinate is the time scale.}  \label{airy_wave_nondamped_front}
\end{figure}

\begin{figure}[h!]
\centerline{
\includegraphics[width=0.6\columnwidth]{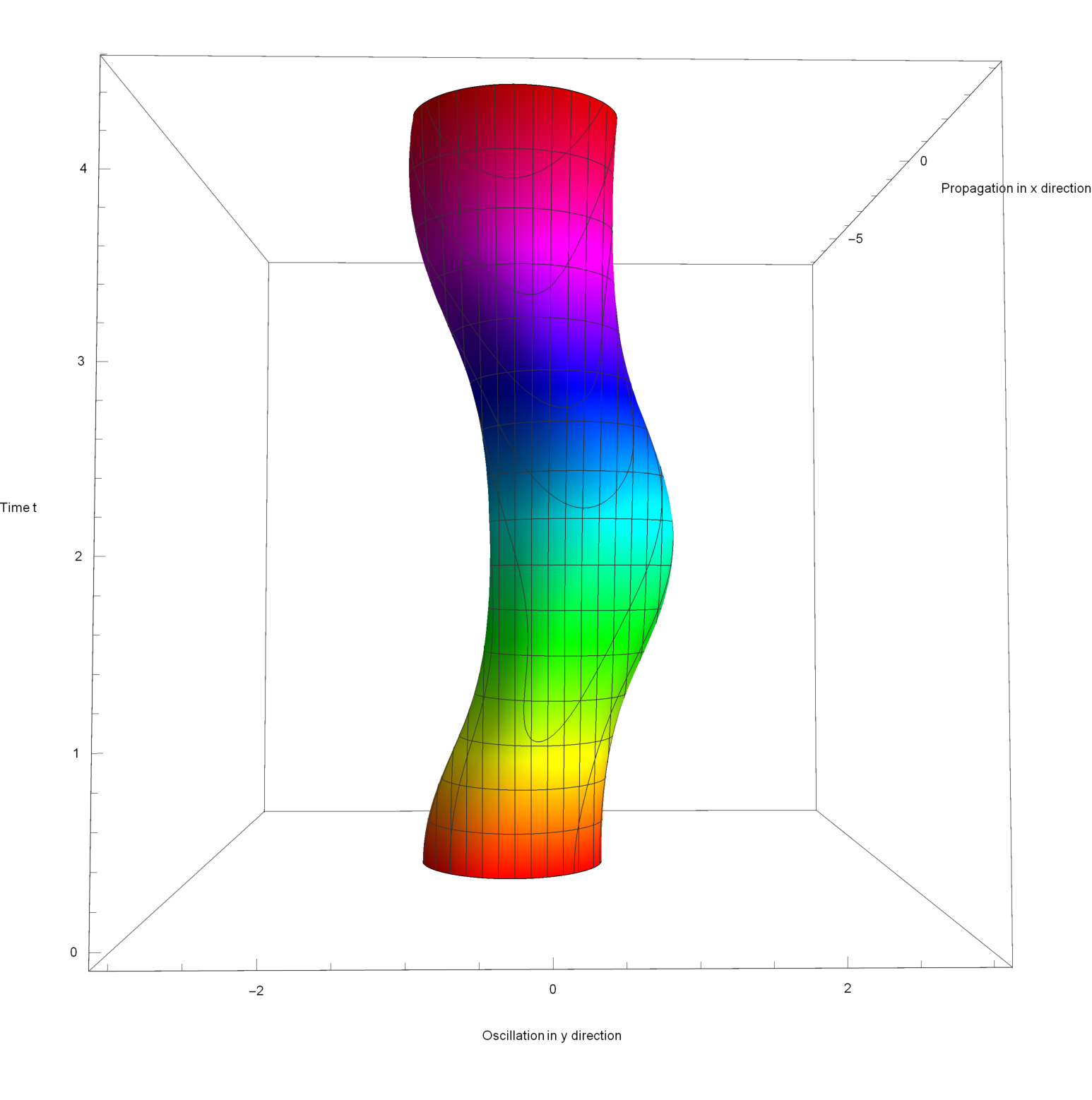}}
\caption{Nondamped oscillating-travelling Airy wave from side view. The translation motion goes opposite us. The perpendicular coordinate is the time scale.}  \label{airy_wave_nondamped_side}
\end{figure}

\begin{figure}[h!]
\centerline{
\includegraphics[width=0.6\columnwidth]{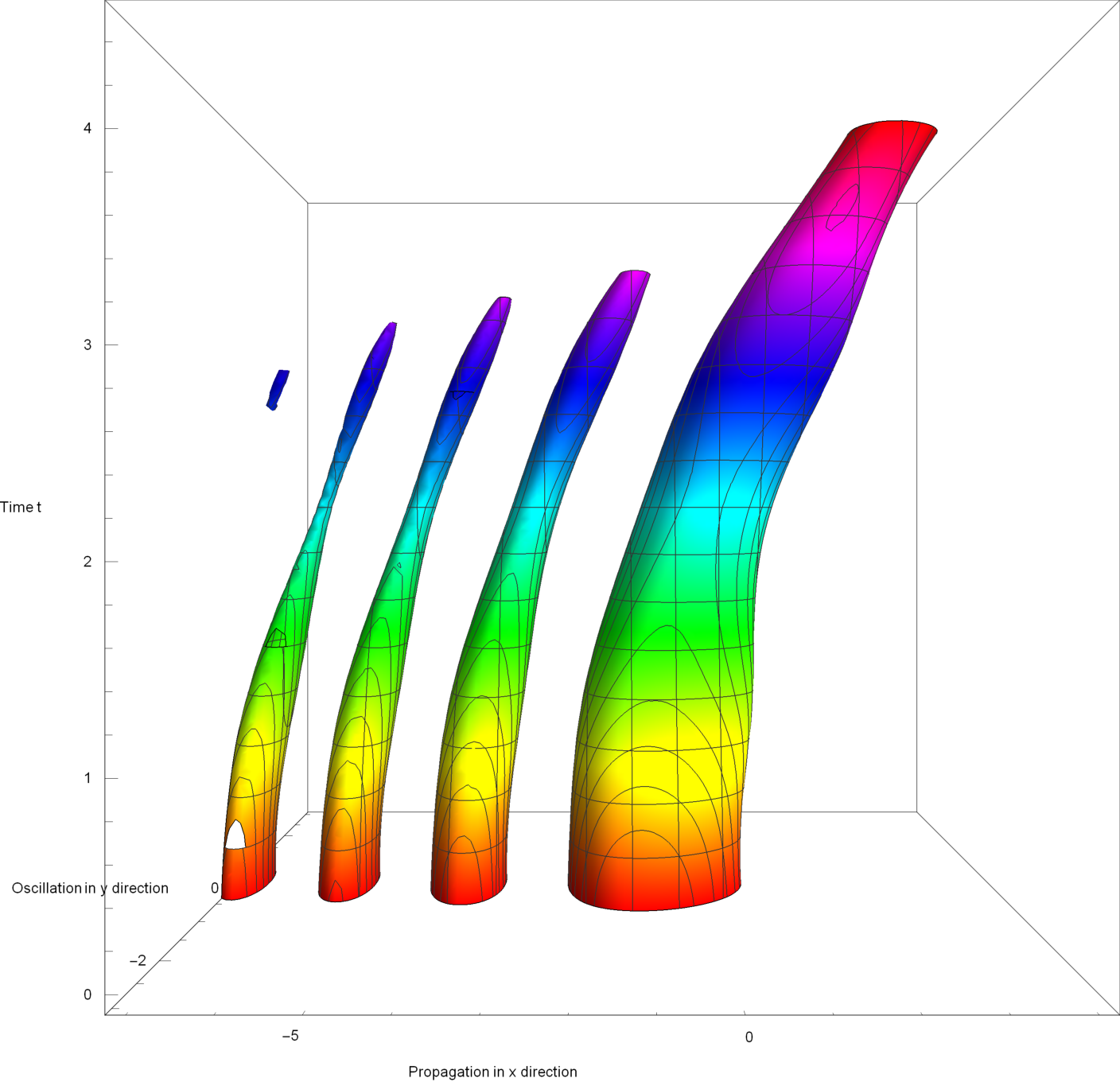}}
\caption{Damped oscillating-travelling Airy wave from front view. The translation
motion is in the x direction, the oscillation is in the y direction. The third coordinate is the
time scale. The dissipation (infromation loss) can be recognized from the decrease of the beam's cross-section.}  \label{airy_wave_damped_front}
\end{figure}

\begin{figure}[h!]
\centerline{
\includegraphics[width=0.6\columnwidth]{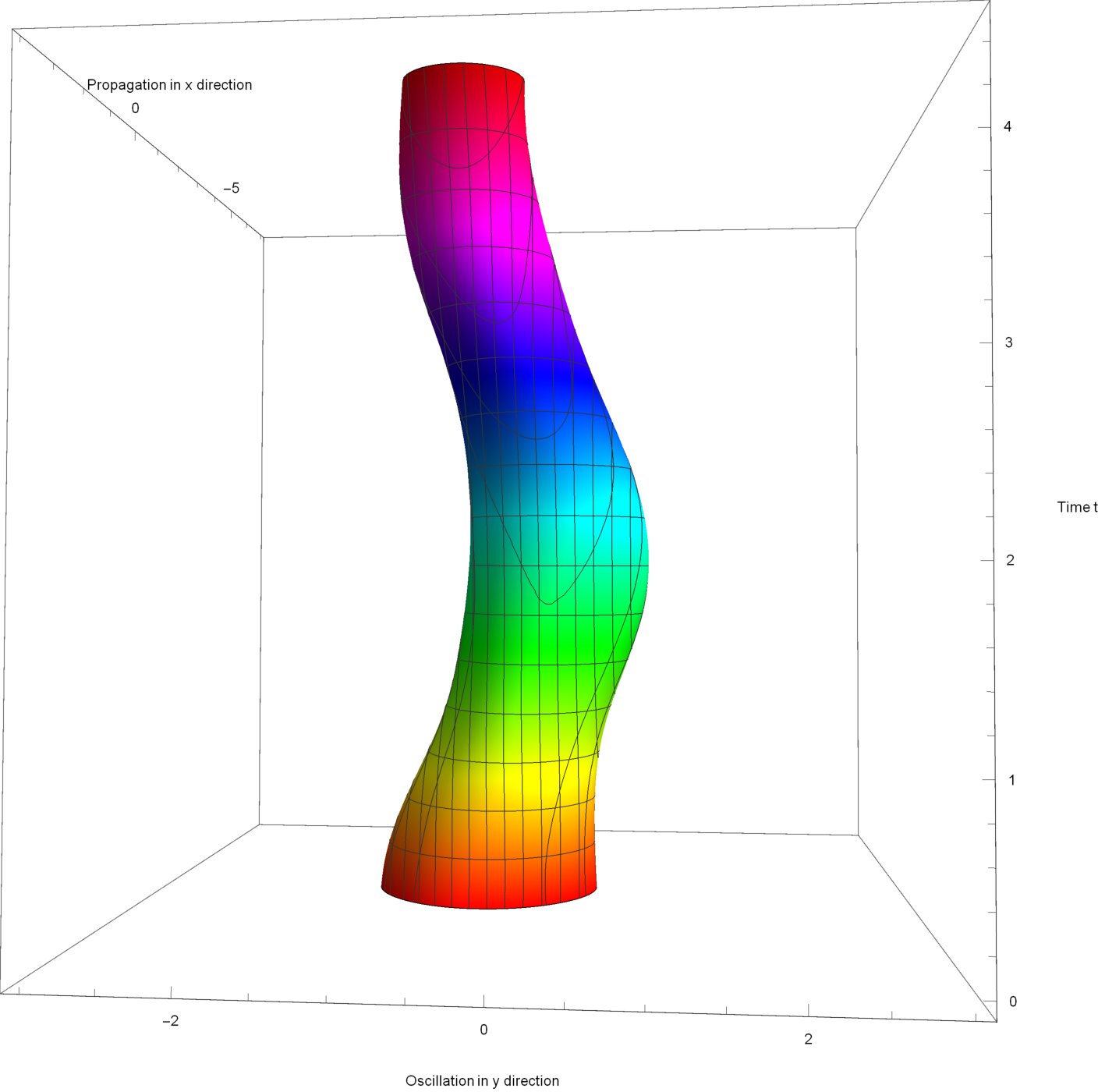}}
\caption{Damped oscillating-travelling Airy wave from side view. The translation motion goes opposite us. The perpendicular coordinate is the time scale. The damping of the oscillation can be recognized well.}  \label{airy_wave_damped_side}
\end{figure}

However, there is another solution of Eq.(\ref{free_propagation_schrodinger}). Berry and Balazs \cite{berry1979} showed the Airy wave train solution of the equation, which describes a diffraction-free, freely accelerating propagation over long distances in the x direction 
\begin{equation}
\psi_{\textrm{Ai}}(x,t) = \textrm{Ai} \left[ \frac{B}{\hbar^{2/3}} \left( x - \frac{B^3 t^2}{4 m^2} \right) \right] \exp \left[ \frac{\mathrm{i} B^3 t}{2 m \hbar} \left( x - \frac{B^2 t^3}{6 m^2} \right) \right].
\end{equation}
For the first look, it seemed only as a mathematical solution possibility. Siviloglou {\it et al}. reported the first observation  of Airy beam \cite{siviloglou2007}. Since then there are more experimental evidence of the existence of the Airy solution \cite{kondakci2017,rozenman2023}. Now, we formulate an alternative solution of the dissipative state equation in Eq. (\ref{dissipative_state_equation})
\begin{eqnarray}
|\Psi(y,x,t)|^2 &=& \frac{1}{\sigma_{y}(t)\sqrt{2}} \exp{ \left\{ -\frac{\left[y - y_{0} \cos(\omega t) \right]^2}{2\sigma_{y}^{2}(y,t)} \right\}} \times \exp \left( -4\lambda t \right) \nonumber \\  &\times&
\textrm{Ai}^2 \left[ \frac{B}{\hbar^{2/3}} \left( x - \frac{B^3 t^2}{4 m^2} \right) \right].  
\end{eqnarray}
The Airy propagating oscillation is damping, but the there is no spreading; it remains diffraction-free. In spite of the amplitude decrease during the propagation the signal can be easily reconstructed. It seems a promising opportunity in the information transfer. We follow the propagating waves in Figs. \ref{airy_wave_nondamped_front} -- \ref{airy_wave_damped_side}. The nondamped propagation is visible from two different views. The typical Airy beam can be identified from the front view in Fig. \ref{airy_wave_nondamped_front}. As it can be recognized, the shape of the beam preserves during the time evolution. The oscillation of the beam can be seen from the side view in Fig. \ref{airy_wave_nondamped_side}. Due to covering, we see opposite us the wavefront. 
%
%
\newline
The damping of the Airy wave packet is obvious, but there is no spreading out as Figs. \ref{airy_wave_damped_front} (front view) and \ref{airy_wave_damped_side} (side view) show. We can continuously identify the remained-structure of the beam. It means the information is always recognizable during the transfer. The damping itself is not equal to the total information loss. The attenuated amplitude can be amplified again. So, finally, the original information can be recovered.

%

The signal reconstruction is of fundamental importance in the information transfer. In the spin relaxation based processes the Loschmidt echo seems promising \cite{csosz2020,csosz2020_2}. The Loschmidt echo is such an effect in which the original state can be recovered quite well.  Hopefully, combining it with our presented ideas, the decrease of the information loss and the state-recovery can be achieved effectively. It is a future challenge.


\section{Conclusion}  \label{conclusion}

The application of potential based canonical quantization procedure is expansible for the damped transversal single particle mechanical waves. The obtained state equation involves the damping in such a way that it pertains to the oscillator and the transversal motion in the same interaction. It reflects the physically relevant situtation that there is only one damping effect in the propagation. The solution of the state equation can be expressed as the product of quantum oscillator, the transversal motion and the exponentially time decreasing factor. The quantum oscillator part relates to the path integral method with the shape-changing formation. There are two opportunity for the transversal solutions. One of these is the Gaussian solution. In this case the signal is spreading, and with the damping together, the carrying information tends to zero. The other possibility is the Airy wave train. Since this solution is non-spreading, diffraction free, shape-reserving, in spite of the damping, the signal is recognizable and repairable for a longer time. The discussed wave propagation mode is promising for the long range quantum signal transfer.  \\ \\




{\bf  Author Contributions:} The authors contributed equally to this work. All authors have read and
agreed to the published version of the manuscript. \\

{\bf Funding:} This research was supported by the National Research, Development and Innovation Office (NKFIH) Grant Nr. K137852 and by the Ministry of Innovation and Technology and the NKFIH within the Quantum Information National Laboratory of Hungary. Supported by the V4-Japan Joint Research Program (BGapEng), financed by the National Research, Development and Innovation Office (NKFIH), under Grant Nr. 2019-2.1.7-ERA-NET-2021-00028. Project no. TKP2021-NVA-16 has been implemented with the support provided by the Ministry of Innovation and Technology of Hungary from the National Research, Development and Innovation Fund. \\

{\bf Data Availability Statement:} Not applicable. \\

{\bf Conflicts of Interest:} The authors declare no conflicts of interest.

\end{document}